\documentclass{ws-p8-50x6-00}
\renewcommand{\rightnote}{{\it Submitted to  {\bf World Scientific} on
November 30, 2000}}%
\begin{document}

\title{Fusion rates in nuclear plasmas}

\author{M.~Coraddu, M.~Lissia and G.~Mezzorani}

\address{
Dipartimento di Fisica, Universit\`a di Cagliari, 
and Istituto
Nazionale di Fisica Nucleare, Sezione di Cagliari,
I-09042 Monserrato, Italy}
\author{P.~Quarati}
\address{
Dipartimento di Fisica, Politecnico di Torino, 
I-10129 Torino, Italy \\
Istituto Nazionale di Fisica Nucleare, Sezione di Cagliari}
\maketitle

\abstracts{
Energy and momentum of the elementary excitations become
independent variables in medium: energy and momentum statistical
distributions are not identical. 
The momentum distribution and not
the energy distribution is relevant for barrier penetration.
The deviations of the momentum distribution from the Maxwell-Boltzmann
energy distribution can be expressed in terms of
the imaginary part of the self-energy of the quasi-particle. It is possible
to obtain an effective Tsallis' distribution for the kinetic energy.
These effects are different from static or dynamical screening and
can have important consequences for reaction rates in stars.}

\section{Introduction}
Theoretical calculations of nuclear rates as function of temperature,
density and composition are fundamental ingredients of our
understanding of stellar structure. Solar standard models 
(SSMs) are based on
rates of the nuclear reactions inside the Sun calculated according
to what is the actual experimental and theoretical
state-of-the-art~\cite{Adel98}.

The solar structure is quite robust and even large changes of
nuclear rates yield non standard solar models~\cite{Castel97} whose structure
is very similar to the standard one; only recent precise measurements
of helioseismic frequencies can discriminate between standard and
non standard models~\cite{helio}.
In addition, there exist changes in chemical
composition, such as those produced by He$^3$ mixing~\cite{He399}, that
cannot be detected even by seismic measurements.

The flux and energy spectrum of neutrinos from the Sun are observables
quite sensitive to some of the reaction rates, but the possibility of
oscillations~\cite{} weakens the link between the observed fluxes and
the rates inside the Sun~\cite{neutrini}.

It is, therefore, of the outmost importance that nuclear rates inside
the Sun, and more in general inside the stars, be accurately calculated
and all possible effects be taken into account.

Cross sections are fundamental for the determination of the rates:
the astrophysical S-factors used in stellar model calculations are
often extrapolations from experimental data at higher energies than
those relevant for stellar interiors; some reactions, {\em e.g.},
the $p+p$ reaction, cannot be measured. Although the underlying
theory is generally robust, new calculations are still necessary.

As an example, a recent and more accurate calculation~\cite{hep} of 
the He$^3 + p$ reaction predicts a S-factor five time larger than
the one used in the SSM. Even if this reaction remains not important for
the basic structure of the Sun, it becomes not negligible for the
high-energy part of the neutrino spectrum. In particular, it could
be important in the interpretation of the possible excess of solar-neutrino
events above 13 MeV, even if the latest experimental data seem to 
suggest that the excess was a statistical fluctuation. Nevertheless,
there exist a vacuum oscillation solution to solar neutrino
problem that reproduces the apparent seasonal variation of the temporal 
series of the GALLEX and Homestake data and that would imply a 
high energy distortion of the
spectrum of the recoil electrons~\cite{excess00}. The most distinct signature
of this solution is a semi-annual seasonal variation of the $^7$Be neutrino
flux with maximal amplitude, future detectors (BOREXINO, LENS and probably
GNO) will be able to test it: the accurate calculation~\cite{hep} of 
the He$^3 + p$ reaction will be very important also for the theoretical
interpretation of these results.

However, nuclear rates depend not only on the reaction cross
sections but also on the
properties of the nuclear plasma. Nuclear plasmas, {\em i.e.},
neutral systems of charged particles (ions and electrons), are
complicated many-body systems: charged are screened, but the
screening depends in general on the energy of the particle and the
remaining interaction is long range and non local. The resulting
spatial and temporal correlations between ions have large effects
especially on fusion reactions, which occur between those high-energy
ions that
can tunnel the Coulomb barrier. 

An accurate theoretical determination
of these rates requires a good understanding of all possible plasma
effects; at the same time, reactions that select ions in the
tail of the energy distribution are probes of the 
dynamics of the plasma itself.

In this contribution we only want to discuss one such plasma effect:
the possibility that the {\em effective momentum distribution} of the
fusing ions could {\em deviate from the standard Maxwell-Boltzmann (MB)
distribution}. In particular, we want to report on our {\em work in progress}
that tries and {\em link the deviation from the MB distribution to the
imaginary part of the self-energy of the quasi-particle states}.

While this effect is potentially important for many astrophysical
phenomena (the solar neutrino problem, brown dwarves, dark matter
distribution), we shall focus on parameters relevant to the solar
core.

\section{Non ideal plasmas}

The solar core is a weakly-nonideal plasma where: 
1) the mean Coulomb energy potential is of the order of the thermal 
kinetic energy; 2) the Debye screening length $R_D\approx a$ (interparticle 
distance): Debye-H\"uckel conditions are only approximately 
verified; 3) it is not possible to separate individual and 
collective degrees of freedom; 4) the inverse solar plasma 
frequency ($t_{pl}=\omega^{-1}_{pl}=
\sqrt{m/4\pi n e^2}\approx 10^{-17}$)
is of the same order of magnitude of the collision 
time $t_{coll}=f^{-1}=\langle n \sigma v\rangle$; 5) particles 
loose memory of the initial state only after 
many collisions: the scattering process cannot be considered Markovian; 
6) the time needed to build up again the screening, after hard collisions, 
is not negligible~\cite{kllq98,ckllmq99}.

At the thermal equilibrium reacting ions are usually described 
as quasi free particles with Maxwell-Boltzmann (MB) velocity distribution.
But many-body effects inside the plasma
could cause deviations from a pure Maxwell-Boltzmann statistics for
the effective degrees of freedom.
Because reacting ions belong to the high momentum tail of the
distribution, at least for fusion reactions between charged ions,
even tiny deviation from the MB tail can cause large modifications
(enhancement or depletion) of the rates.

The value of the collision frequency $f$ determines the possibility
of two different
effects that produce important deviations 
from the  Maxwellian distribution $F_M(p)$  at high momenta:
\begin{itemize}
\item[$Q$)]
Quantum uncertainty effect.\\
When the Coulomb collisional frequency is large ($h f>kT$) the ions
cannot be considered as quasi-free particles: the energy and momentum
distributions are different and one must decide which one is relevant
for the reaction rates. The fact that the two distributions are not
equivalent is related to the finite life-time of the 
quasi-particles and to the quantum uncertainty. Since
nuclear rates should be evaluated averaging the quasi-classical cross 
section $\sigma(p)$ over the momentum distribution, rather 
than the energy distribution, even if the energy distribution is
Maxwellian, the effective distribution can acquire a non-Maxwellian 
tail~\cite{Sav99,Star00}.
\item[$q$)]
Weak nonextensivity effect.\\
Tsal\-lis statistics~\cite{Tsallis} with entropic parameter $q$
can describe systems that are not extensive due to long-range interactions 
or non-Markovian memory effects; the energy distribution itself
deviate from the standard free-particle statistics. When deviations are
small ($q\approx 1$) the correction (enhanced or depleted tail) can
be described  by the factor 
$\exp\left [-\frac{1-q}{2}\left(\frac{\epsilon_p}{kT}\right)^2\right]$.
\end{itemize}

Deviations from the Maxwellian tail due to either $Q$ or $q$ effect
(or both) may
lead to strong increase or decrease of the nuclear rates in the 
solar core (non standard solar models due to large changes of the
nuclear rates and their implications for the solar neutrino problem
are described in Ref.~\cite{Castel97}).

In this contribution we discuss only the $Q$ effect, {\em i.e.}, the
deviation of the momentum distribution relative to the energy
distribution. As we shall see, this effect leads only to an
enhanced tail.

\section{Quasi-particle momentum distribution}

Many properties of interacting systems can often be described by weakly
interacting excitations or quasi-particles. The energy-momentum dispersion
relation (position of the pole of the one-particle Green's function) of
these excitations is  found by solving
\begin{equation}
\omega = \frac{p^2}{2m} + \Sigma(\omega, p^2) \quad,
\end{equation}
where $\Sigma(\omega, p^2)= \Sigma_R + i \Sigma_I $ is the self-energy of
the one-particle propagator.

In the approximation of a constant real part of the self-energy $\Sigma_R$,
we obtain the shift of the energy due to the static mean-field (in
plasma it produces non-dynamical screening).

The $p^2$ dependence of $\Sigma_R$ reflects the spatial
nonlocality of the effective interaction and may be understood qualitatively
by considering the nonlocality of the exchange term of the Hartree-Fock
potential, while the $\omega$ dependence reflects the nonlocality of $\Sigma$
in time.

As long as the imaginary part $\Sigma_I$ is zero, there exist a one-to-one
correspondence between the energy  $\omega$ and momentum
$p$ (or kinetic energy $p^2/(2m)$). In real systems the imaginary part
of the self-energy of the quasi-particle is not zero and energy and
momentum become independent variables; however, they are still strongly
correlated when $\Sigma_I$ is small (only if $\Sigma_I$ is small
the concept of quasi-particle is useful).

Barrier penetration is determined by the momentum of the (quasi-)par\-ti\-cle
and not by its energy (when they do not coincide).

In this preliminary presentation, we shall restrict ourselves
to the case of an  energy distribution that is Maxwellian: 
$P(E)\sim \exp(-\beta E)$.

If we are given the relation between $E$ and $p^2$ in the form
$F(E, p^2)$, the momentum distribution is
obtained 
\begin{equation}
P(p^2) = \int_0^{\infty} dE e^{-\beta E} F(E, p^2) \quad .
\end{equation}

For free particles $F(E, p^2) = \delta\left( E-p^2/(2m)\right)$ and,
therefore,
\begin{equation}
P(p^2) = e^{-\beta\frac{ p^2}{2m} } \quad .
\end{equation}

If $\Sigma_I=0$ and $\Sigma_R$ is constant, then
$F(E, p^2) = \delta\left( E - p^2/(2m) - \Sigma_R\right)$ and the
distribution is still Maxwellian
\begin{equation}
P(p^2) = e^{-\beta\left( \frac{ p^2}{2m}+\Sigma_R \right) } \quad .
\end{equation}

If $\Sigma_R$ is not constant but can be expanded in the region of interest
\begin{equation}
\Sigma_R(\omega,p^2) = \Sigma_R +
\frac{\partial \Sigma_R}{\partial p^2} (p^2 - p_0^2) + 
\frac{\partial \Sigma_R}{\partial \omega} (\omega - \omega_0) 
\end{equation}
the result is
\begin{equation}
P(p^2) = e^{-\beta\left( \frac{ p^2}{2m^*}+\Sigma_R \right) } \quad ,
\end{equation}
where 
\begin{equation}
m^*= m \left( 1+ 2 m \frac{\partial\Sigma_R}{\partial p^2}\right)^{-1} 
\left(1- \frac{\partial\Sigma_R}{\partial \omega} \right) \quad .
\end{equation}

In all the above cases the distribution for the variable $p^2$ is still
Maxwellian (in general it follows the energy distribution), even if the
shift in energy or the effective mass can have
important phenomenological consequences (screening, level densities, etc.).

When the imaginary part of the self-energy cannot be disregarded 
($\Sigma_I > 0 $), there appear deviations from the Maxwellian 
distribution. For the sake of discussion let us consider the following
relation between $\omega$ and $\epsilon_p \equiv p^2 / (2m^*)+\Sigma_R$:
\begin{equation}
\label{FnuwTsallis}
F_{\nu}(\omega, p^2) = \frac{1}{\Gamma(1/\nu)} \frac{1}{E}
           \left( \frac{E}{\nu \epsilon_p}\right)^{\frac{1}{\nu}}
           e^{-\frac{E}{\nu \epsilon_p}} \quad ,
\end{equation}
where the parameter $\nu$ characterizes the deviation from
a $\delta$-function.
This function is normalized
\begin{equation}
\int_0^{\infty} dE F_{\nu}(E, p^2) =
 \frac{1}{\Gamma(1/\nu)} \int_0^{\infty} dx x^{1/\nu-1} e^{-x} = 1 \quad ,
\end{equation}
and in the limit $\nu\to 0^+$
\begin{equation}
\log(F_{\nu})
   =  \frac{1}{\nu}\left[ 1 -\frac{E}{\epsilon_p}  
  + \log\frac{E}{\epsilon_p} \right]  - \frac{1}{2}\log(2\pi\nu E^2) + O(\nu)
   \label{smlnu}
\end{equation}
The function between bracket $f(x) = 1-x+\log(x)\leq 0$ and is zero
only for $x=1$; therefore, if $E\neq \epsilon_p$,  $\lim_{\nu\to 0^+}
 \log(F_{\nu}) = -\infty$ and $\lim_{\nu\to 0^+} F_{\nu} = 0$. If
$E = \epsilon_p$, $\lim_{\nu\to 0^+} F_{\nu} = 
  \frac{1}{\sqrt{2\pi\nu} E} = \infty $. In summary
\begin{equation}
\lim_{\nu\to 0^+} F_{\nu}(E,p^2)  = \delta\left( E-\epsilon_p\right) \quad .
\end{equation}
If $\nu$ is small but not zero, we try the {\em Ansatz} $E/\epsilon_p =
1 + a \sqrt{\nu}$ 
into the equation~(\ref{smlnu}) and find
\begin{equation}
\log(F_{\nu})
       = - \frac{a^2}{2} - \frac{1}{2}\log(2\pi\nu E^2) + O(\nu)
\label{gauexp}
\end{equation}
where the terms that have been dropped are really of order $\nu$
as long as $a = (E/\epsilon_p - 1)/\sqrt{\nu}$ remains of order one,
{\em i.e.}, as long as $E/\epsilon_p - 1$ does not becomes large compared
to $\sqrt{\nu}$ (remains of order $\sqrt{\nu}$).

From the definition $a^2 = (E/\epsilon_p-1)^2 / \nu $ 
we can rewrite Eq.~(\ref{gauexp})
\begin{equation}
\lim_{\nu\to 0^+} F_{\nu} =  \frac{1}{\sqrt{2\pi\nu}E} 
       e^{ -\frac{1}{2}
       \left( \frac{E-\epsilon_p}{\sqrt{\nu}\epsilon_p} \right)^2 }
          =  \frac{1}{\sqrt{2\pi\nu}\epsilon_p} 
       e^{ -\frac{1}{2}
       \left( \frac{E-\epsilon_p}{\sqrt{\nu}\epsilon_p} \right)^2 } \quad .
\end{equation}

Since $\sqrt{\nu}\epsilon_p$ is the width of the distribution and
the imaginary part of the self-energy $\Sigma_I$ is also proportional to
the width of the distribution (when it is in Lorenzian form),
it is plausible  that
\begin{equation}
\nu = C \left( \frac{\Sigma_I}{\epsilon_p} \right)^2 \quad ,
\end{equation}
at least in the limit $\Sigma_I \ll \epsilon_p$; the constant $C$ depends
on the precise definition of the limit of the quasi-particle
(we are working on the microscopical derivation of such kind of
relation and of $C$).

If we instead calculate the distribution of $\epsilon_p$ from the
relation $F_{\nu}(\omega,p^2)$ between energy and momentum for general
$\nu$ (without expansion for small $\nu$), we find
\begin{eqnarray}
\int_0^{\infty}d\omega e^{-\beta \omega} F(\omega,p^2) &=&
\int_0^{\infty}d\omega e^{-\beta \omega} 
\frac{1}{\Gamma(1/\nu)} \frac{1}{\omega}
           \left( \frac{\omega}{\nu \epsilon_p}\right)^{\frac{1}{\nu}}
           e^{-\frac{E}{\nu \epsilon_p}} \nonumber\\
&=& 
\left( 1+ \nu \beta\epsilon_p \right)^{-\frac{1}{\nu}} \quad,
\end{eqnarray}
which is the Tsallis' distribution with $\nu = q-1\geq 0$.

It is not possible to obtain a Tsallis' distribution with $\nu = q-1 < 0$
by this kind of effect, since the broadening of dispersion relation
between energy and momentum physically has the effect of increasing the
tail of the distribution: it cannot ``cut'' the tail.

At least in the limit of small $\Sigma_I$, we have given a phenomenological
interpretation of the parameter $q$ in the Tsallis' distribution
\begin{equation}
q-1 = C \left( \frac{\Sigma_I}{\epsilon_p} \right)^2 \quad .
\end{equation}

\section{Width comparison}
We are studying a more general approach that would allow
to compare different relations between $p^2$ and $E$, at least in
the asymptotic limit that
each relation $F(E,p^2)$ tends to a $\delta$-function,
$F(E,p^2)\to \delta(E-\epsilon_p)$. As long as we consider
one-parameter generalizations of the $\delta$-function,
there should be some mapping from one parameterization to the other.
We have already seen how the relation in  Eq.~(\ref{FnuwTsallis}) 
and a Gaussian relation coincide in the limit of small width. 

An other interesting case is the Lorenzian,
which also becomes a $\delta$-function in the limit of vanishing
width.
In fact, the  quasi-particles dispersion relation becomes in dense
media (at least in the limit of large life-time)~\cite{Galit67}
\begin{equation}
\delta_g(\epsilon)=\frac{1}{\pi} 
\frac{g(\epsilon,p)}{[(\epsilon-\epsilon_p-
\Delta(\epsilon,\epsilon_p))^2+g(\epsilon,\epsilon_p)^2]}\; ,
\end{equation}
where $\epsilon_p=p^2/2m$, 
$\Delta(\epsilon,\epsilon_p)$ and $g(\epsilon,p)$ are the real 
and ima\-gi\-na\-ry parts of the one-particle retarded Green's 
function self-energy.

For weakly non ideal plasmas one can show that 
$\Delta\approx kT\, \Gamma/2$, where $\Gamma$ is the pla\-sma parameter 
($\Gamma=e^2/R_D kT$) and $g \propto h f$. At non zero value of $g$, 
a nonexponential tail appears in the distribution function $F_Q(p)$.  
For large momenta, it has been found~\cite{Star98a,Star98b} that
\begin{equation}
F_Q(p) =
F_M(p)+\frac{h f}{2\pi} \,\frac{kT}{\epsilon_p^2} \,e^{\mu/kT} \; ,
\end{equation}
where $\mu$ is the chemical potential and $F_M(p)$ the Maxwellian distribution.

\section{Conclusions}
We have shown that even when ions have a MB energy
distribution the finite life-time of the quasi-particles in the plasma
can produce a non-Maxwellian momentum distribution. Since the tunneling
probability between charged ions must be evaluated using the momentum
distribution, the reaction rates are effectively obtained by using
distributions that depart from the MB  one. This departure from the
MB distributions can be calculated if one knows the spectral
dispersion relation between energy and momentum. 

In particular, we have shown that it is possible to have relations
between energy and momentum that yield momentum distributions of
Tsallis type with entropic parameter $q>1$, which corresponds to an
enhanced tail.

In this framework, we have suggested a possible interpretation of
the parameter $q$ in terms of the imaginary part of the quasi-particle
self-energy, at least in the limit of $q\to 1$ (small deviations
from MB).

\section*{Acknowledgments}
It is a pleasure to thank the Organizers for the very interesting
and fruitful Meeting.
This work was supported by ``Confinanziamento MURST-PRIN''.

\end{document}